\documentclass[twocolumn,prd,groupedaddress,nofootinbib]{revtex4}

\usepackage{graphicx}
\usepackage{dcolumn}
\usepackage{bm}

\begin{document}

\title{
QCD/String holographic mapping and high energy scattering amplitudes}

\author{Henrique Boschi-Filho}
\email{boschi@if.ufrj.br}
\affiliation{Instituto de F\'{\i}sica, 
Universidade Federal do Rio de Janeiro, Caixa Postal 68528, RJ 21941-972 
-- Brazil}
\author{Nelson R. F. Braga}
\email{braga@if.ufrj.br}
\affiliation{Instituto de F\'{\i}sica,
Universidade Federal do Rio de Janeiro, Caixa Postal 68528, RJ 21941-972 
-- Brazil}

 
\begin{abstract}
We find a one to one mapping between low energy string dilaton states
in AdS bulk and high energy glueball states on the corresponding boundary. 
This holographic mapping leads to a relation between bulk and boundary 
scattering amplitudes. From this relation and the dilaton action we find 
the appropriate momentum scaling for high energy QCD amplitudes at fixed 
angles.  
\end{abstract}

\maketitle


\vfill\eject

The idea that strong interactions should have a description in terms 
of strings is not new \cite{Planar} (see \cite{Pol} for a discussion). 
However, understanding the non trivial relation between string theory 
and QCD is still a defying task.  
A remarkable step in this direction was given by Maldacena \cite{Malda}, 
conjecturing the equivalence of large $N$ limit of $SU(N)$ 
superconformal field theories and 
supergravity/string theory in a higher dimensional anti-de 
Sitter spacetime (AdS/CFT correspondence) \cite{GKP,Wi,Malda2}.  
In this correspondence, bulk fields act as classical sources for boundary 
correlation functions\cite{GKP,Wi,Malda2,MV,FMMR}.
One of the striking features of this  correspondence, pointed out soon by 
Witten \cite{Wi}, is that it is a realization of the holographic principle: 
``The degrees of freedom of  a quantum theory with gravity can be mapped 
on the corresponding boundary"\cite{HOL1,HOL2,HOL3,HOL4,HOL5}. 
From general arguments of the AdS/CFT correspondence one finds that there 
is an isomorphism between the Hilbert spaces of bulk and boundary  
theories \cite{HS1,HS2,HS3,HS4}. 

A long standing puzzle 
for the string description of strong interactions, 
is the high energy scattering at fixed angles.
In string theory such a process is soft (the amplitudes decay 
exponentially with energy) while both experimental data and QCD 
theoretical predictions\cite{QCD1,BRO} indicate a hard behavior 
(amplitudes decaying with a power of energy).
A solution for this puzzle was  proposed recently by 
Polchinski and Strassler\cite{PS} based on the AdS/CFT scenario. 
They introduced an energy scale (associated with the lightest 
glueball mass) by cutting the AdS space and taking a slice 
analogous to that of the Randall-Sundrum model \cite{RS1,RS2}. 
This AdS slice is considered as an approximation for the space 
dual to a confining gauge theory which can be associated with QCD.
Then they found the correct high energy QCD amplitude for glueball 
scattering at fixed angles starting from the string amplitude in 10 
dimensions and integrating over the warped AdS extra dimension, 
weighted by the dilaton wave function.

A general relation between string and QCD states 
would certainly be non trivial. 
However, for the particular process of glueball 
high energy scattering at fixed angles 
the result of Polchinski and Strassler 
seems to suggest that the states of 
strings and QCD could be related explicitly. 
At low energies string theory can be approximated by a
supergravity action involving the dilaton field.
In this regime one can look for a mapping between states 
of the dilaton and of boundary fields.

In this letter we consider the same kind of AdS slice of Polchinski 
and Strassler \cite{PS} and propose a one to one mapping between 
bulk low energy string dilaton states and boundary composite 
operators associated with QCD glueballs.
This mapping is inspired in a simpler version found in ref.\cite{BB2} 
between Fock spaces of  scalar field theories in  AdS bulk and boundary.
Here we consider glueball operators  and map their scattering amplitude 
into the bulk dilaton amplitude.
Note that bulk and boundary operators should be on-shell in order
to represent in and out asymptotic states of scattering processes. 
Then using the low energy string action and restricting to fixed angle 
scattering (in both theories) we find that the glueball amplitudes scale 
as
\begin{equation}
\label{scale}
{\cal M} \,\,\sim \,\, s^{(4-\Delta )/2}
\end{equation}

\noindent where $\sqrt{s}$ is the energy and
 $\Delta$ is the sum of the scaling dimensions of glueball operators. 
This is the  expected behavior for high energy QCD\cite{QCD1,BRO}
and is in agreement with Polchinski and Strassler \cite{PS}.

Let us consider a type IIB string theory. At energies much lower than 
the string scale $1/\sqrt{\alpha^\prime}\,$ this theory can be described  
by  the supergravity action\cite{GKP,Po}
\begin{equation}
S \,=\, {1\over 2 \kappa^2} \int d^{10}x \sqrt{G} e^{-2\Phi} \Big[
{\cal R} + G^{MN} \partial_M \Phi \partial_N \Phi  \,+\, ....\Big]
\end{equation}

\noindent 
where $G^{MN}$ is the ten dimensional metric, ${\cal R}$ is the Ricci 
scalar curvature,  $\Phi$ is the dilaton field and 
$\kappa \sim g (\alpha^\prime)^2 $. 
We  identify this ten dimensional space as AdS$_{5}\,\times S^5$
with radius $R$ and measure
\begin{equation}
\label{metric}
ds^2=\frac {R^2 }{ z^2}\Big( dz^2 \,+(d\vec x)^2\,
- dt^2 \Big) \,+ \,R^2 d\Omega_5^2 \,\,,
 \end{equation}

\noindent 
where the coordinates $\Omega_5$ describe a five dimensional sphere
$S^5$.  

Following Polchinski and Strassler \cite{PS} we cut the AdS space 
in order to obtain a boundary theory with an energy scale (mass gap).
We consider an AdS slice with "size" $z_{max}$ representing an 
infrared cut off associated with the mass of the lightest glueball.  
We also consider the dilaton to be in the $s-wave$ state, 
so we will not take 
into account variations with respect to $S^5$ coordinates. 
Thus, the action becomes 

\begin{eqnarray}
\label{Action}
S &=& {\pi^3 R^8 \over 4 \kappa^2} \int d^4x \int_0^{z_{max}} {dz\over z^3} 
 e^{-2\Phi}\nonumber\\
 &\times& \Big[
{\cal R} + ( \partial_z \Phi )^2 \,+\, \eta^{\mu\nu} \partial_\mu \Phi  
\partial_\nu \Phi\,+\, ....\Big]
\end{eqnarray}

\noindent where $\eta^{\mu\nu}$ is the four dimensional Minkowiski metric.
 
The solution for the free dilaton field, that will be used to build up 
the bulk asymptotic states in the scattering process can be cast 
into the form \cite{BB1}
\begin{eqnarray}
\label{QF}
\Phi(z,\vec x,t) &=& \sum_{p=1}^\infty \,
\int { d^3 k \over (2\pi)^{3}}\,
{z^{2} \,J_2 (u_p z ) \over z_{max} w_p(\vec k ) 
\,J_{3} (u_p z_{max} ) }\nonumber\\
&\times& \lbrace { {\bf a}_p(\vec k )\ }
 e^{-iw_p(\vec k ) t +i\vec k \cdot \vec x}\,
\,+\,\,h.c.\rbrace\,,
\end{eqnarray}

\noindent with $0\,\le\,z\,\le z_{max}\,$
and $w_p(\vec k ) \,=\,\sqrt{ u_p^2\,+\,{\vec k}^2}\,$ , $h.c.\,$ 
means hermitean conjugate and  
$u_p$ are defined by 
\begin{equation}
\label{up}
u_p z_{max}\,=\, \chi_{_{2\,,\,p}}
\end{equation}

\noindent 
such that the Bessel function satisfies $ J_2 (\chi_{_{2\,,\,p}} )=0$. 
If we had considered bulk fields with mass $M$ the order of this Bessel 
function would be $\,\nu = \sqrt{4 + M^2 R^2}\,$. The operators 
${\bf a}_p\, ,\;{\bf a}^{\dagger}_p \,$ satisfy the commutation relations
\begin{equation}
\label{canonical1}
\Big[ {\bf a}_p(\vec k )\,,\,{\bf a}^\dagger_{p^\prime}({\vec k}^\prime  )
\Big]\,=\, 2\, (2\pi)^{3} w_p(\vec k )   
\delta_{p\,  p^\prime}\,\delta^{3} (\vec k -
{\vec k}^\prime )\,.
\end{equation}
 
\noindent In the AdS/CFT correspondence bulk scalars couple to 
composite conformal boundary operators of dimension 
$d \,=\, 2 + \sqrt{4 + M^2R^2}$.

On the boundary ($ z = 0)$ of the AdS slice we consider massive 
composite operators $\Theta(\vec x,t)\,$ that will represent 
glueballs. 
The corresponding creation-annihilation operators are assumed to 
satisfy the algebra (for asymptotic states) 
\begin{equation}
\label{canonical2}
\Big[ {\bf b}( \vec K )\,,\,{\bf b}^\dagger ({ \vec K }^\prime  )
\Big]\,=\, 2 (2\pi)^{3} \, w( \vec K ) \,\delta^3 ( \vec K -
{ \vec K}^\prime ) \,,
\end{equation}

\noindent where $ w(\vec K ) = \sqrt{ {\vec K}^2 + \mu^2}$. 
Note that this commutation relation holds only for
asymptotic states. In general, 
the creation-annihilation operators associated with a
composite field like $\Theta(\vec x,t)\,$ cannot satisfy 
such a free field relation.

A mapping between theories that live in different dimensions is not 
trivial. 
In the standard AdS/CFT framework there is a correspondence between  
(on shell) string theory in the AdS bulk and 
(off shell) conformal field theory on the boundary. 
Here we are considering a different and simpler situation of 
string theory in a low energy regime where it can be approximated 
by a supergravity action. 
In particular, we are interested in the dilaton field which 
is the bulk dual to the boundary glueball operator. 
So we are looking for a mapping between two field 
theories defined in different dimensions. 
Further, we want to relate asymptotic states of bulk and boundary 
theories so that we will consider (on shell) quantized free fields. 
If both field theories had continuous momenta it would be impossible
to find a one to one mapping between quantum states. 
However, as we consider just a slice of AdS, 
the spectrum of the momentum associated with the axial 
direction is discrete. 
Actually even when one considers not just a slice but the whole AdS 
space one must include a boundary at infinity, compactifying the space 
and finding again a discrete spectrum in the axial direction, 
as discussed in \cite{BB1}.
Then the continuous part of bulk and boundary momenta 
$\vec k$ and $\vec K$ have the same dimensionality. 

This discretization makes it possible to establish 
a one to one mapping between bulk
and boundary momenta $(\vec k , u_p)$ and $\vec K$. 
We assume a trivial mapping between their angular parts
and then look for a relation between  their moduli
$K \equiv \vert \vec K \vert \, , \,k \equiv \vert \vec k \vert \,$.
The idea is that one can relate the points of an enumerable set of 
$p\,$ lines with those of a single line by dividing the latter into 
intervals and mapping each line $\,p\,$ into a corresponding interval.
So we introduce a sequence of energy scales ${\cal E}_1 \,,\,
{\cal E}_2\,,{\cal E}_3\,...$ 
in order to map each interval of the boundary momentum modulus
${\cal E}_{p-1} < K \le {\cal E}_p \,$ with $p=1,2,...$  
into the entire range of the transverse bulk momentum modulus $ k$,  
corresponding to some fixed axial momentum $u_p$. 
This mapping between bulk and boundary momenta allow us to
map the corresponding creation-annihilation operators 
(\ref{canonical1},\ref{canonical2}).
For the first energy interval, corresponding to $u_1$,  defined as 
$0 \le K \le {\cal E}_1\,$, we can write\cite{BB2} 
\begin{eqnarray}
\label{ab}
k\,{\bf a}_1 ( \vec  k ) 
&=& K \,{\bf b}( \vec K  ) \nonumber\\
k\,{\bf a}^\dagger_1 ( \vec k ) 
&=& K\,{\bf b}^\dagger ( \vec K  )\,.
\end{eqnarray}

\noindent This mapping must preserve the physical consistency
of both theories. In particular Poincare invariance should not be broken
neither for the boundary theory nor for the bulk theory at a fixed $z$.
This is obtained imposing that the canonical commutation relations 
(\ref{canonical1},\ref{canonical2}) are preserved by the mapping (\ref{ab}).
Then substituting eq. (\ref{ab}) into relation (\ref{canonical1}) and using 
eq. (\ref{canonical2}) 
we find an equation in terms of the moduli of bulk and boundary
momenta which solution can be written as\cite{BB2}
\begin{equation}
\label{completo}
k = {u_1 \over 2} 
\,\Big[ \,{ {\cal E}_1  +\sqrt{{\cal E}_1^2 + \mu^2 } 
			\over  K + \sqrt{K^2 + \mu^2}}
- { K + \sqrt{K^2 + \mu^2}\over {\cal E}_1  
+\sqrt{{\cal E}_1^2 + \mu^2 }}\,\Big]\,.
\end{equation}

\noindent Similar relations can be obtained for the other intervals 
${\cal E}_{p-1} < K \le {\cal E}_p \,$ with $p=2,3,...$.
One might wonder if the trivial mapping $\vec k = \vec K$ would also 
be a solution.
However this would not provide a one to one mapping between the  
entire bulk momenta
$(\vec k , u_p )$ and  
the boundary momenta $\vec K$ for all different values of $p$
as long as there is only one boundary field with mass $\mu$. 

The momentum operators in the bulk and boundary theories are respectively:
\begin{eqnarray} 
(\vec P ,u) &=& \sum_p \int {d^3k \over 2 (2\pi)^3 } 
{{\bf a}^\dagger_p ( \vec k ) {\bf a}_p ( \vec k ) \over \sqrt{k^2 + u_p^2}}
 (\vec k , u_p )\\
\vec \Pi &=&  \int {d^3K \over 2 (2\pi)^3 } 
{{\bf b}^\dagger ( \vec K ) {\bf b} ( \vec K ) \over \sqrt{K^2 + \mu^2}}
 \vec K
\end{eqnarray}
\noindent Note that Poincare invariance in the $\vec x$ directions holds 
both in 
the boundary and bulk theories since the canonical commutation relations 
(\ref{canonical1}) and (\ref{canonical2}) are preserved by the mapping.
 
We are considering a high energy glueball scattering  
on the four dimensional AdS boundary. 
This  will be mapped into a scattering process
of dilaton states in the effective low energy string theory.
The equation (\ref{completo}) 
is understood as the relation between bulk and 
boundary momenta for the particles involved in the scatterings.
Identifying $\mu$ with
the mass of the lightest glueball and choosing the AdS size as  
\begin{equation}
z_{max} \,\sim\, {1\over \mu}\,
\end{equation}
 
\noindent we find that $u_1 \sim \mu $, once $z_{max} \,\sim 1/u_1$
according to eq. (\ref{up}). Note that the size $z_{max}$ 
can then be interpreted as
an infrared cutoff for the boundary theory.

Further, we can take  ${\cal E}_1\,$ large enough so that the 
momenta associated with the high energy glueball scattering can 
fit into the region $\mu \ll K \ll {\cal E}_1$. 
Then we can  approximate relation (\ref{completo}) as
\begin{equation}
\label{Kk}
k \,\,\approx\,\, { {\cal E}\, \mu  \over 2 \, K}\,, 
\end{equation}

\noindent where we defined ${\cal E}_1\,\equiv\,{\cal E}\,$ and 
disregarded the other energy intervals
associated with higher axial momenta $u_p\,,\,p\ge\,2\,$.
Note that this mapping together with the conditions 
$\mu \ll K \ll {\cal E} $ imply  that $\mu \ll k \ll {\cal E} $.

Choosing the string scale to  be of the same order of the high 
energy cut off of the boundary theory, i.e., 
${\cal E}\,\sim 1/ \sqrt{\alpha^\prime}$,
we find that the momenta $k$ associated with string theory 
correspond to a low energy
regime well described by the supergravity approximation (\ref{Action}).

The equations (\ref{ab}) and (\ref{Kk})  represent a one to one 
holographic mapping
between bulk dilaton  and boundary glueball states.
Now we are going to use these equations to relate the corresponding 
scattering amplitudes.

Let us consider, in the bulk string theory, a scattering  of
2 particles in the initial state and $m$ particles in the final state,
with all particles having axial  momentum  $u_1$. The $S$ matrix reads 
\begin{eqnarray} 
S_{Bulk} &=&  \langle \, {\vec k}_3\,,u_1;\,...; \,{\vec k}_{m+2}\,,u_1\,
;\,out \vert {\vec k}_1,\,u_1\, 
;\,{\vec k}_2 \,,u_1;in\,\rangle \nonumber\\
&=&  \langle \, 0\,\vert \,{\bf a}_{out} ( {\vec k}_3 )\,... \,{\bf a}_{out}
 ({\vec k}_{m+2}) \,  {\bf a}^+_{in }
( {\vec k}_1) \,{\bf a}^+_{in} ({\vec k}_2) \,\vert \, 0\, \rangle 
\,,\nonumber\\
& &
\end{eqnarray}

\noindent where ${\bf a} \equiv {\bf a}_1$ and the $in$ and $out$ 
states are defined as 
$ \vert \vec k \,,\,u_1\,\rangle 
\,=\, {\bf a}^+ (\vec k ) \vert 0 \rangle \,$.

Now using the mapping between creation-annihilation operators 
(\ref{ab}) one can rewrite the above $S$ matrix in terms of 
boundary operators.
Considering fixed angle scattering, we take the bulk momenta 
to be of the form
$ k_i \,=\, \gamma_i k$ and the boundary momenta 
$ K_i \,=\, \Gamma_i K$, where
$\gamma_i $ and $\Gamma_i$ are constants with $i\,=\,1,2,...,m+2\,$. 
Then
\begin{widetext}
\begin{eqnarray}
S_{Bulk}&\sim &  \langle  0 \vert \,{\bf b}_{out} ( {\vec K}_3 )\,... \,
\,{\bf b}_{out} ({\vec K}_{m+2})   
\,{\bf b}^+_{in }( {\vec K}_1)\, {\bf b}^+_{in} ({\vec K}_2) \vert 0 \rangle  
\,\Big({ K \over k}\Big)^{m + 2 } \nonumber\\
&\sim& \, \langle  \,{\vec K}_3 \,,... \,{\vec K}_{m+2},\,out \,\vert \,{\vec K}_1 \,,
{\vec K}_2 \,,in\,\rangle \, \Big( { K \over k} \Big)^{m+ 2} K^{(m+2)(d-1) }\,,
\end{eqnarray}
\end{widetext}
\noindent where the composite operators on the boundary have  scaling dimension $d$
and then their $in$ and $out$ states are 
$ \vert \vec K \,\rangle \,\cong\, K^{ 1 - d } {\bf b}^+ (\vec K ) \vert 0 \rangle \,,$
within the regime $K \gg \mu$.

Using the relation (\ref{Kk}) between bulk and boundary momenta we get
\begin{equation}
S_{Bulk} \, \sim \,  
 S_{Bound.} \,\,\Big( {\sqrt{\alpha^\prime} \over \mu }\Big)^{m+2} \,\, K^{(m+2)(1+d)}
\end{equation}

As the scattering amplitudes ${\cal M}$ are related to  the corresponding $S$ matrices 
(for non equal $in$ and $out$ states) by
\begin{eqnarray}
S_{Bulk} &=& {\cal M}_{Bulk} \, \delta^4 (  k_1^\rho +  k_2^\rho -  k_3^\rho - \,...\,- 
 k_{m+2}^\rho )
\nonumber\\
 S_{Bound.} &=& {\cal M}_{Bound.} \, \delta^4 ( K_1^\rho +  K_2^\rho 
- K_3^\rho - ... -K_{m+2}^\rho  )\nonumber\\
& &
\end{eqnarray}

\noindent we find a relation between bulk and boundary scattering amplitudes
\begin{eqnarray}
\label{Mb}
{\cal M}_{Bound.} &\sim& {\cal M}_{Bulk}\,\,  S_{Bound.}\,\, (\, S_{Bulk}\,)^{-1} 
\Big( { K\over k} \Big)^4 \nonumber\\
&\sim&  {\cal M}_{Bulk}\, \,K^{8 -  (m+2)(d + 1)  } \,\,\,
\Big( {\sqrt{\alpha^\prime} \over \mu }\Big)^{2 - m}\,\,.
\end{eqnarray}

Now we must evaluate the bulk amplitude from the string low energy 
effective action (\ref{Action}). 
The momentum dependence of this amplitude can be determined using dimensional arguments.
Note that the global constant 
$\, R^8 / \kappa^2\, $ associated with this action is dimensionless
and the only dimensionfull parameters are $z_{max}\,\,\sim\,1/\mu\,$ and the 
Ricci scalar $ {\cal R} \,\sim \, 1/R^2\,$. 
As $\mu\,\ll\,k$ the relevant contribution to the bulk amplitude will not involve
$z_{max}$.  Further, choosing the condition  $ 1/R \ll k $
we can disregard the term involving the Ricci scalar.
This condition does not fix completely the AdS radius $R$ and we additionally 
impose that $\mu \ll 1/R\,$. This implies that $z_{max} \gg R$.
Then, if one regularizes the divergence ($z = 0$) of the bulk action by cutting the axial
coordinate
$z$ at  $R$ as in \cite{GKP}, one still has a large portion of
the original AdS space: $ R \,\le \,z \,\le \,z_{max}\,$.
This guarantees that we keep the interesting AdS region which is 
an approximation for the near horizon geometry of $N$ coincident $D3$-branes,
as in the Maldacena duality.

Taking into account the normalization of the states $\vert k, u_1\rangle\,$
one sees that $\,{\cal M}_{Bulk}\,$ is dimensionally [Energy]$^{4 - n}$,  where $n$ 
is the total number of scattered particles. As $k$ is the only dimensionfull
quantity that is relevant at leading order for the bulk scattering in the 
regime considered we find:
\begin{equation}
{\cal M}_{Bulk}\,\sim\, k^{ 2 - m }\,\,.
\end{equation}

\noindent Using again the relation between bulk and boundary 
momenta (\ref{Kk}) and inserting this result in the boundary amplitude (\ref{Mb})
we get
\begin{equation}
{\cal M}_{Boundary} \,\sim \,K^{4 - \Delta } \,,
\end{equation}

\noindent where $\Delta = ( m + 2) d $ is the total scaling dimension
of the scattering particles associated with glueballs on the four 
dimensional boundary. Considering  $ K \sim \sqrt{ s} $ we find 
the expected QCD scaling behavior\cite{QCD1,BRO}
\begin{equation}
 {\cal M}_{Boundary} \,\sim \,s^{2  - \Delta/2 } 
\,.
\end{equation}

\noindent This shows that the bulk/boundary one to one mapping (\ref{ab}) , (\ref{Kk})
can be used to obtain the hard scattering behavior of high energy glueballs 
at fixed angles, from a low energy approximation of string theory. 

It is interesting to relate the different energy scales used in the above
derivation of the scattering amplitudes and check their consistency. 
The scales we discussed are 
\begin{equation}
\mu \,\,\ll \,\,{1\over R}\,\,\ll\,\, {1\over \sqrt{\alpha^\prime}}\,.
\end{equation}

\noindent Note that the relation between the AdS radius $R$ , the number of
coincident branes $N$, the string coupling constant $g$ and scale $\alpha^\prime$
is $R^4 \,\sim \, g N ( \alpha^\prime )^2$.
Then the above condition between $R$ and $\alpha^\prime$ corresponds
to the $\,$ 't Hooft limit\cite{Planar}.
Assuming that the dimensionless quantity $\mu R$ is the parameter
that relates the energy scales we find $ \sqrt{\alpha^\prime}\,=\,\mu R^2 \,$
so that the lightest glueball mass is
\begin{equation}
\mu^2 \,=\,{1\over g N \alpha^\prime}\,\,.
\end{equation}

\noindent  This result is in agreement with \cite{MN}. Further,  the 
above relation between the energy scales together with the condition that $k \gg 1/R\,$
and the mapping between bulk and boundary momenta (\ref{Kk}) imply that
\begin{equation}
\mu \,\,\ll K \,\ll \,{1\over R}\,\ll\,k\,\ll\,\, {1\over \sqrt{\alpha^\prime}}\,\,,
\end{equation}

\noindent so that the absolute values of $k$ are greater than those of $K$,
although the boundary scattering is a high energy process
(with respect to $\mu$) while the bulk scattering is a low  energy
process (with respect to $\,1/\sqrt{\alpha^\prime}\,$).
Furthermore  $K$ and $k$ in this regime are inversely proportional
showing a kind of infrared-ultraviolet duality as expected from  holography.
We hope that the kind of mapping discussed in this paper might 
be extended to other physical processes or energy regimes.
In particular we have recently applied this mapping to obtain the scalar 
glueball mass spectrum\cite{BB3}.

\section*{Acknowledgments}
We would like to thank Nathan Berkovits for discussions. 
The authors are partially supported by CNPq, FINEP , 
CAPES (Procad program) and FAPERJ 
- Brazilian research agencies.



\end{document}